\documentclass[11pt,letterpaper]{article}

\usepackage{typearea}
\typearea{15} 

\usepackage{shadow}
\usepackage{theorem,latexsym,graphicx,amssymb}
\usepackage{amsmath,enumerate}
\usepackage{wrapfig}
\usepackage{epsfig}
\usepackage{xspace}
\usepackage{bm}
\usepackage{layout}
\usepackage{shadethm}
\usepackage[compact]{titlesec}
\usepackage{algorithm, algorithmic}
\usepackage{color}
\usepackage{ifpdf}

\definecolor{Darkblue}{rgb}{0,0,0.4}
\definecolor{Brown}{cmyk}{0,0.81,1.,0.60}
\definecolor{Purple}{cmyk}{0.45,0.86,0,0}
\newcommand{\mydriver}{hypertex}
\ifpdf
 \renewcommand{\mydriver}{pdftex}
\fi
\usepackage[breaklinks,\mydriver]{hyperref}
\hypersetup{colorlinks=true,%pdfborder={1 1 1 [3]},%
            citebordercolor={.6 .6 .6},linkbordercolor={.6 .6 .6},%
citecolor=Darkblue,urlcolor=black,linkcolor=Darkblue,pagecolor=black}
\newcommand{\lref}[2][]{\hyperref[#2]{#1~\ref*{#2}}}


\newcommand{\ignore}[1]{}

\newcommand{\R}[0]{{\ensuremath{\mathbb{R}}}}

\newcommand{\Z}[0]{{\ensuremath{\mathbb{Z}}}}

\newcommand{\poly}{\operatorname{poly}}

\newcommand{\norm}[2]{\ensuremath{\left\Vert {#1} \right\Vert_{#2}}}

\newcommand{\bigvec}[1]{\ensuremath{\bigl\langle #1 \bigr\rangle}}

\newcommand{\class} [1] {\ensuremath{{\bf{#1}}}\xspace}
\renewcommand{\P} {\class{P}}
\newcommand{\NP} {\class{NP}}

\newcommand{\cost}{\ensuremath{{\sf fac}}}
\newcommand{\sse}{\subseteq}

\newcommand{\OPT}{\ensuremath{\underline{\sf opt}}}
\newcommand{\ALG}{\ensuremath{\underline{\sf alg}}}
\newcommand{\fhat}{\widehat{f}}
\newcommand{\rhat}{\widehat{f}}
\newcommand{\g}{f^*}

\newcommand{\junk}[1]{}

\newcommand{\ts}{\textstyle}

\newcounter{note}[section]

\newtheorem{theorem}{Theorem}[section]

\newtheorem{lemma}[theorem]{Lemma}
\newshadetheorem{lemmashaded}[theorem]{Lemma}
\newtheorem{fact}[theorem]{Fact}
\newtheorem{claim}[theorem]{Claim}

\newenvironment{proof}{

\noindent{\bf Proof:}}
{\hfill$\blacksquare$

}

\newcommand{\initOneLiners}{%
    \setlength{\itemsep}{0pt}
    \setlength{\parsep }{0pt}
    \setlength{\topsep }{0pt}
}
\newenvironment{OneLiners}[1][\ensuremath{\bullet}]
    {\begin{list}
        {#1}
        {\initOneLiners}}
    {\end{list}}

\newcommand{\kmedcf}{\ensuremath{{\sf kmed}}\xspace}
\newcommand{\uflcf}{\ensuremath{{\sf UFL}}\xspace}
\newcommand{\kufl}{\ensuremath{{\sf kUFL}}\xspace}
\title{Simpler Analyses of Local Search 
  Algorithms\\ for Facility Location\thanks{Computer Science Department, Carnegie Mellon 
    University, Pittsburgh, PA 15213. Research was partly supported by
    the NSF awards CCF-0448095 and CCF-0729022, and an Alfred P.~Sloan Fellowship.}}
\author{
Anupam Gupta
\and
Kanat Tangwongsan}

\begin{document}

%\begin{titlepage}
%\def\thepage{}
%\thispagestyle{empty}

\date{}
\maketitle

% Draft: notice
% ======================
% \begin{center}
% \vspace{-1cm}
% \bf{\huge{DRAFT}}
% \end{center} 
% ======================

\begin{abstract}

  \medskip\noindent We study local search algorithms for metric
  instances of facility location problems: the uncapacitated facility
  location problem (UFL), as well as uncapacitated versions of the
  $k$-median, $k$-center and $k$-means problems. All these problems
  admit natural local search heuristics: for example, in the UFL problem
  the natural moves are to open a new facility, close an existing
  facility, and to swap a closed facility for an open one; in
  $k$-medians, we are allowed only swap moves. The local-search
  algorithm for $k$-median was analyzed by Arya et al.~(\emph{SIAM
    J.~Comput.~33(3):544-562, 2004}), who used a clever ``coupling''
  argument to show that local optima had cost at most constant times the
  global optimum. They also used this argument to show that the local
  search algorithm for UFL was $3$-approximation; their techniques have
  since been applied to other facility location problems.
  
  \medskip\noindent In this paper, we give a proof of the $k$-median
  result which avoids this coupling argument. These arguments can be
  used in other settings where the Arya et al.\ arguments have been
  used. We also show that for the problem of opening $k$ facilities $F$
  to minimize the objective function $\Phi_p(F) = \big(\sum_{j \in V}
  d(j, F)^p\big)^{1/p}$, the natural swap-based local-search algorithm
  is a $\Theta(p)$-approximation. This implies constant-factor
  approximations for $k$-medians (when $p=1$), and $k$-means (when $p =
  2$), and an $O(\log n)$-approximation algorithm for the $k$-center
  problem (which is essentially $p = \log n$).
\end{abstract}

%\end{titlepage}

%\newpage
%======================================================
% intro
%======================================================
\section{Introduction}
\label{sec:introduction}

Facility location problems have been central objects of study in the
operations research and computer science community, not only for their
intrinsic fundamental nature and broad applicability, but also as
problems whose solutions have led to the development of new ideas and
techniques. Indeed, techniques such as rounding linear relaxations,
using primal-dual techniques and Lagrangean relaxations, greedy
algorithms (with and without the dual-fitting approach), and local
search algorithms have all been honed when applied to facility
location problems.

Local search has been a popular algorithm design paradigm, and has had
many successes in the design of approximation algorithms for hard
combinatorial optimization problems. The focus of this paper is on local
search algorithms for facility location problems on metric spaces.  In
fact, the best approximation algorithm known for the $k$-median problem
is a $(3 + \varepsilon)$-approximation via local search~\cite{AGKMMP01}.
However, the analysis of this simple local search algorithm is fairly
subtle, and requires a careful ``coupling'' argument. The same coupling
argument was used by~\cite{AGKMMP01} to also analyze a local search
algorithm for uncapacitated facility location, and subsequently
by~\cite{DGKPSV05, Pandit-thesis} for some other problems.

In this paper, we present the following:
\begin{itemize}
\item We give somewhat simpler analyses for the natural local search
  algorithms of~\cite{AGKMMP01,DGKPSV05,Pandit-thesis}: while the
  approximation guarantees remain the same, the proofs are arguably more
  intuitive than existing proofs.
  
\item We show that the problem of opening $k$ facilities $F$ to minimize
  the objective function
  \[ \Phi_p(F) = \bigg(\sum_{j \in V} d(j, F)^p\bigg)^{1/p}, \] the
  natural swap-based local-search algorithm is a
  $\Theta(p)$-approximation\footnote{When we say that our local search
    algorithm is a $\rho$-approximation, we mean that the cost of every
    local optimum is at most $\rho$ times the optimum cost. In this
    paper, we also implicitly mean that one can find a solution of cost
    $(\rho+\varepsilon)OPT$ in time $\poly(n, \varepsilon)$---see
    \lref[Section]{sec:related-work} for details.}: this immediately
  implies constant factor approximations for $k$-medians (which
  corresponds to the case $k=1$), $k$-means (the case $p = 2$), an
  $O(\log n)$-approximation local search algorithm for the $k$-center
  problem (which is essentially the case $p = \log n$).  To the best of
  our knowledge, the results for $p \neq 1$ give the first analyses of
  local search heuristics for these problems.
\end{itemize}
% We hope that the simpler analyses presented in this paper will lead to a
% better understanding of local search algorithms for facility location
% problems, and will lead to new results.

% \agnote{Is this the first constant factor for $k$-means in general
%   metrics?}

\paragraph{Technical Ideas.} The main contribution of this work is the
simplification of the proofs. As with most local search proofs, the
previous papers considered a local optimum, and showed that since a
carefully chosen set of local moves were non-improving, we could infer
some relationship between our cost and the optimal cost. However, in the
previous papers, this set of local moves have to be carefully defined by
looking at how the clients served by each optimal facility were split up
between facilities in the local optimum. Even bounding the change in
cost due to each of these potential local moves is somewhat non-trivial.

In our paper, we define a set of local moves based only on distance
information about which of the optimal facilities $F^*$ are close to
which of our facilities $F$. The intuition is simple: consider each of
the optimal facilities in $F^*$, and look at the closest facility to
it in $F$. If some facility $f \in F$ is the closest to only one
facility $f^* \in F^*$, then we should try swapping $f$ with
$f^*$. However, if there is some facility $f \in F$ that is the
closest facility to many facilities in $F^*$, then swapping $f$ might
be bad for our solution, and hence we do not want to close this
facility in any potential move. Formalizing this natural intuition
gives us the claimed simpler proofs for many problems.

\subsection{A Note on the Rate of Convergence.} 
\label{sec:note-rate-conv}
In this paper, we will only focus on the quality of the local optima,
and not explicitly deal with the rate of convergence. This, however, is
only for brevity: since all our arguments are based on averaging
arguments, we can show that if at any point in the local search
procedure, the current solution is very far from every local optimum,
then there is a step that reduces the objective function by a large
amount. Moreover, since we are dealing with approximation algorithms, we
can stop when each local step improves the objective function by at most
a factor of $(1+\varepsilon)$, which would give us only a slightly worse
approximation guarantee but would keep the running time polynomial in
$n$ and $\varepsilon^{-1}$. These details are fairly standard; e.g., see
the discussion in~\cite{AGKMMP01}.

\subsection{Other Related Work}
\label{sec:related-work}

Facility location problems have had a long history; here, we mention
only some of the results for these problems. We focus on metric
instances of these problems: the non-metric cases are usually much
harder~\cite{Hoch82}.

\textbf{$k$-Median.} The $k$-median problem seeks to find facilities
$F$ with $|F| = k$ to minimize $\sum_{j \in V} d(j, F)^2$. The first
constant factor approximation for the k-median problem was given by
Charikar et al.~\cite{CGTS02}, which was subsequently improved
by~\cite{CG99} and \cite{AGKMMP01} to the current best factor of $3 +
\varepsilon$. It is known that the natural LP relaxation for the
problem has an integrality gap of~$3$, but the currently-known
algorithm that achieves this does not run in polynomial
time~\cite{ARS03}. The extension of $k$-median to the case when one
can open at most $k$ facilities, but also has to pay their facility
opening cost was studied by~\cite{DGKPSV05}, who gave a
$5$-approximation.

\textbf{$k$-Means.} The $k$-means problem minimizes $\sum_{j \in V}
d(j, F)^2$, and is widely used for clustering in machine learning,
especially when the point set is in the Euclidean space. For Euclidean
instances, one can obtain $(1 + \varepsilon)$-approximations in linear
time, if one imagines $k$ and $\varepsilon$ to be constants: see
\cite{KSS04} and the references therein.  The most commonly used
algorithm in practice is Lloyd's algorithm, which is a local-search
procedure different from ours, and which is a special case of the EM
algorithm~\cite{Lloyd}.  While there is no explicit mention of an
approximation algorithm with provable guarantees for $k$-means (to the
best of our knowledge), many of the constant-factor approximations for
$k$-median can be extended to the $k$-means problem as well. The paper
of Kanungo et al.\cite{KMNPSW03} is closely related to ours: it
analyzes the same local search algorithm we consider, and uses
properties of k-means in Euclidean spaces to obtain a
$9$-approximation.  Our results for hold for general metrics, and can
essentially be viewed as extensions of their results.

\textbf{$k$-Center.} Tight bounds for the $k$-center problem are known:
there is a $2$-approximation algorithm due to
\cite{Gonzalez-kcenter,HochShm86}, and this is tight unless $\P = \NP$.

\textbf{UFL.} For the uncapacitated metric facility location (UFL)
problem, the first constant factor approximation was given by Shmoys et
al.~\cite{STA97}; subsequent approximation algorithms and hardness
results have been given by~\cite{STA97, Chudak98,Sviridenko02, ChuShm03,
  Byrka07, JV99,CG99,PT03, MMSV01, JMS02-greedyFL, MYZ02-FL,
  KPR98,CG99,AGKMMP01, GuhaKhuller}. It remains a tantalizing problem to
close the gap between the best known approximation factor of
$1.5$~\cite{Byrka07}, and the hardness result of
$1.463$~\cite{GuhaKhuller}.

% The first constant factor approximation for the $k$-median problem was
% given by Charikar et al.~\cite{CGTS02}, which was subsequently improved
% by~\cite{CG99} and~\cite{AGKMMP01} to the current best factor of $3 +
% \varepsilon$. The best hardness result known is a $1 + 2/e$-hardness.
% The extension of this problem to the case when one could open at most
% $k$ facilities, but also had to pay their facility opening cost was
% studied by~\cite{DGKPSV05}, who gave a $5$-approximation.

% Tight bounds for the $k$-center problem are known: there is a
% $2$-approximation algorithm due to~\cite{Gonzalez-kcenter,HochShm86},
% and this is tight unless $\PCLASS=\NP$.

% The $k$-means problem minimizes $\sum_{j \in V} d(j,F)^2$, and is widely
% used for clustering in machine learning, especially when the point set
% is in Euclidean space. For Euclidean instances, one can obtain
% $(1+\varepsilon)$-approximations in linear time, if one imagines $k$ and
% $\epsilon$ to be constants: see~\cite{KSS04} and the references therein.
% The most commonly used algorithm in practice is Lloyd's algorithm, which
% is an instance of the EM algorithm. To the best of our knowledge, there
% is no constant factor approximation known for general metric spaces for
% the $k$-means problem.

\subsection{Organization of the Paper}
\label{sec:org}
The paper is organized as follows. In \lref[Section]{sec:basic-kM}, we
consider the metric $k$-Median problem and present an analysis showing
a $5$ approximation.  Following that, we analyze a version of the
local-search algorithm for $k$-Median that allows $t$ simultaneous
swaps, where we show a $(3 + 2/t)$ approximation. In
\lref[Section]{sec:lp-kfl}, we consider a common generalization of the
$k$-median, $k$-means, and $k$-center problems, called the
$\ell_p$-norm $k$-facility location problem, where we give an $O(p)$
approximation algorithm, and present an instance which is
asymptotically tight.  Finally, in \lref[Section]{sec:grab-bag}, we
present simpler proofs of two other known results: a $3$ approximation
for the uncapacitated facility location problem, and a $5$
approximation for the $k$-uncapacitated facility location problem.

\section{A Simpler Analysis of $k$-Median Local Search}
\label{sec:basic-kM}

In this section, we study the local search algorithm for the
$k$-median problem, and show that any local optimum has a cost which
is at most $5$ times the cost of the global optimum. Recall that given
a set $F$ of at most $k$ facilities, the $k$-median cost is
\[ \kmedcf(F) = \sum_{j \in V} d(j, F). \] 

The local search algorithm we consider in this section is the simplest
one: we start with any set of $k$ facilities. At each point in time,
we try to find some facility in our current set of facilities, and
swap it with some currently unopened facility, so that the cost of the
resulting solution decreases. It is known that any local minimum is a
$5$-approximation to the global minimum~\cite{AGKMMP01}, and that the
bound of $5$ is tight for these local-search dynamics. Here we give a
simpler proof of this $5$-approximation.

\subsection{A Set of Test Swaps}
\label{sec:test-swaps}

To show that a local optimum is a good approximation, the standard
approach is to consider a carefully chosen subset of potential swaps:
if we are at a local optimum, each of these swaps must be
non-improving.  This gives us some information about the cost of the
local optimum. To this end, consider the set $F^*$ of facilities
chosen by an optimum solution, and let $F$ be the facilities at the
local optimum. Without loss of generality, assume that $|F| = |F^*| =
k$.

\begin{figure}[!h]
  \centering
%  \fbox{Figure here}
    \includegraphics[scale=.65]{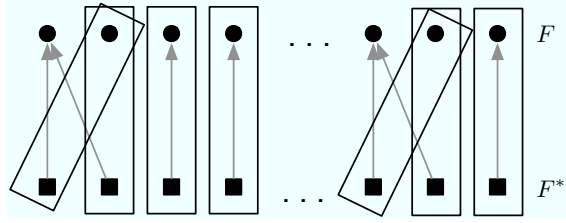}
    \caption{An example mapping $\eta\!: F^* \to F$ and a set of test
      swaps $S$.}
\end{figure}

Define a map $\eta: F^* \to F$ that maps each optimal facility $f^*$ to
a closest facility $\eta(f^*) \in F$: that is, $d(f^*, \eta(f^*)) \leq
d(f^*, f)$ for all $f \in F$. Now define $R \sse F$ to be all the
facilities that have \emph{at most}~$1$ facility in $F^*$ mapped to it
by the map $\eta$.  (In other words, if we create a directed bipartite
graph by drawing an arc from $f^*$ to $\eta(f^*)$, $R \sse F$ are those
facilities whose in-degree is at most~$1$). 

Finally, we define a set of $k$ pairs $S = \{(r,f^*)\} \sse R \times
F^*$ such that
\begin{OneLiners}
\item Each $f^* \in F^*$ appears in exactly one pair $(r,f^*)$.
\item If $\eta^{-1}(r) = \{f^*\}$ then $r$ appears only once in $S$
  as the tuple $(r, f^*)$.
\item If $\eta^{-1}(r) = \emptyset$ then $r$ appears at most in two
  tuples in $S$. 
\end{OneLiners}

The procedure is simple: for each $r \in R$ with in-degree $1$,
construct the pair $(f, \eta^{-1}(r))$---let the optimal facilities that
are already matched off be denoted by $F^*_1$. The other facilities in
$R$ have in-degree $0$: denote them by $R_0$. A simple averaging
argument shows that the unmatched optimal facilities $|F^* \setminus
F^*_1| \leq 2|R_0|$. Now, arbitrarily create pairs by matching each node
in $R_0$ to at most two pairs in $F^* \setminus F^*_1$ so that the above
conditions are satisfied.

The following fact is immediate from the construction:
\begin{fact}
  \label{fct:valid-map}
  For any tuple $(r,f^*) \in S$ and $\fhat^* \in F$ with $\fhat^*
  \neq f^*$, $\eta(\fhat^*) \neq r$.
\end{fact}

\paragraph{Intuition for the Pairing.} 
To get some intuition for why the pairing $S$ was chosen, consider the
case when each facility in $F$ is the closest to a unique facility in
$F^*$, and far away from all other facilities in $F^*$---in this case,
opening facility $f^* \in F^*$ and closing the matched facility in $f
\in F$ can be handled by letting all clients attached to $f$ be handled
by $f^*$ (or by other facilities in $F$). A problem case would be when a
facility $f \in F$ is the closest to several facilities in $F^*$, since
closing $f$ and opening only one of these facilities in $F^*$ might
still cause us to pay too much---hence we never consider the gains due
to closing such ``popular'' facilities, and instead only consider the
swaps that involve facilities from the set of relatively ``unpopular''
facilities $R$.

\subsection{Bounding the Cost of a Local Optimum}

In this section, we use the fact that each of the swaps in set $S$
constructed in \lref[Section]{sec:test-swaps} are non-improving to show
that that the local optimum has small cost.

Breaking ties arbitrarily, assume that $\varphi: V \to F$ and
$\varphi^*: V \to F^*$ are functions mapping each client to some closest
facility.  For any client $j$, let $O_j = d(j,F^*) = d(j,\varphi^*(j))$
be the client $j$'s cost in the optimal solution, and $A_j = d(j,F) =
d(j, \varphi(j))$ be it's cost in the local optimum. Let $N^*(f^*) = \{
j \mid \varphi^*(j) = f^*\}$ be the set of clients assigned to $f^*$ in
the optimal solution, and $N(f) = \{j \mid \varphi(j) = f\}$ be those
assigned to $f$ in the local optimum.

% \begin{figure}[h!]
% \centering
% \includegraphics[scale=.65]{reassign1}
% %\caption{The candidate reassignment for $j \in N(r)\setminus N^*(f^*)$.}
% \end{figure}

\begin{lemma}
  \label{lem:ubd-kmed}
  For each swap $(r,f^*) \in S$, 
  \begin{gather}
    \label{eq:1}
    \kmedcf(F + f^* - r) - \kmedcf(F) \leq \sum_{j \in N^*(f^*)} (O_j - A_j) +
    \sum_{j \in N(r)} 2\,O_j.
  \end{gather}
\end{lemma}

\begin{proof}
  Consider the following candidate assignment of clients (which gives
  us an upper bound on the cost increase): map each client in
  $N^*(f^*)$ to $f^*$. For each client $j \in N(r) \setminus
  N^*(f^*)$, consider the figure below. Let the facility $\fhat^* =
  \varphi^*(j)$: assign $j$ to $\rhat = \eta(\fhat^*)$, the closest
  facility in $F$ to $\fhat^*$.  Note that by
  \lref[Fact]{fct:valid-map}, $\rhat \neq r$, and this is a valid new
  assignment. All other clients in $V \setminus (N(r) \cup N^*(f^*))$
  stay assigned as they were in $\varphi$.
  \begin{center}
    \includegraphics[scale=.65]{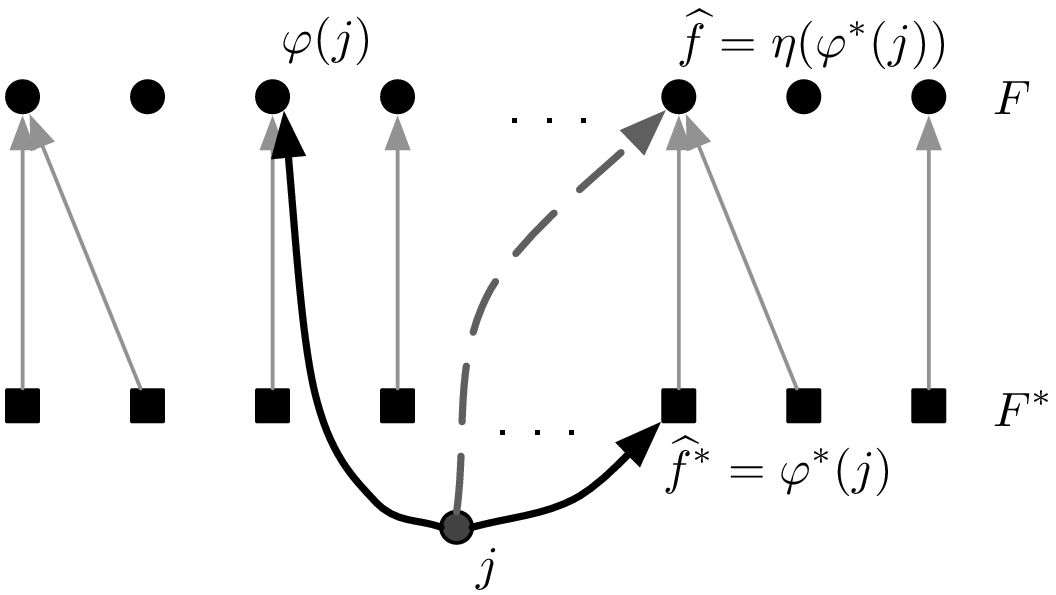}
  \end{center}
  Note that for any client $j \in N^*(f^*)$, the change in cost is
  exactly $O_j - A_j$: summing over all these clients gives us the first
  term in the expression~(\ref{eq:1}).

  For any client $j \in N(r) \setminus N^*(f^*)$, the change in cost is
  \begin{align}
    d(j,\rhat) - d(j,r) 
    &\leq d(j,\fhat^*) + d(\fhat^*, \rhat) - d(j,r) \label{eq:2}\\
    &\leq d(j,\fhat^*) + d(\fhat^*, r) - d(j,r) \label{eq:3}\\
    &\leq d(j,\fhat^*) + d(j, \fhat^*) = 2\,O_j. \label{eq:4}
  \end{align}
  with~(\ref{eq:2}) and~(\ref{eq:4}) following by the triangle
  inequality, and~(\ref{eq:3}) using the fact that $\rhat$ is the
  closest vertex in $F$ to $\fhat^*$. Summing up, the total change for
  all these clients is at most
  \begin{gather}
    \sum_{j \in N(r) \setminus N^*(f^*)} 2\,O_j \leq \sum_{j \in N(r)}
    2\,O_j,
  \end{gather}
  the inequality holding since we are adding in non-negative terms. This
  proves \lref[Lemma]{lem:ubd-kmed}.
\end{proof}

Note that summing~(\ref{eq:1}) over all tuples in $S$, along with the
fact that each $f^* \in F^*$ appears exactly once and each $r \in R \sse
F$ appears at most twice gives us the simple proof of the following
theorem.
\begin{theorem}[\cite{AGKMMP01}]
  \label{thm:localmin-kmed}
  At a local minimum $F$, the cost $\kmedcf(F) \leq 5\cdot \kmedcf(F^*)$.
\end{theorem}

\subsection{A Projection Lemma}

Like the proof for $k$-Median, most proofs in the remaining of the
paper also rely on a candidate assignment of clients to give an upper
bound on the cost increase. In particular, these proofs often reassign
a client $j$ to $\eta(\varphi^*(j))$, the image of the facility
$\varphi^*(j)$ that serves $j$ in the optimal solution. The following
lemma bounds the distance between client $j$ and the facility
$\eta(\varphi^*(j))$; its proof is exactly the same as in
(\ref{eq:2})-(\ref{eq:4}).

%\begin{shadebox}

\begin{lemmashaded}[Projection Lemma]
  \label{lem:prjlem}
  For any client $j \in C$,  
  \[ d(j, \eta(\varphi^*(j)) \leq 2O_j + A_j.
  \]
\end{lemmashaded}

%\end{shadebox}
%\begin{proof} The proof uses the triangle inequality and the fact that
%   $\eta(\varphi^*(j))$ is closer to $\varphi^*(j)$ than any other node
%   in $F$.
%   \begin{align*}
%     d(j, \eta(\varphi^*(j)) 
%     &\leq
%     d(j, \varphi^*(j)) +    d(\varphi^*(j), \eta(\varphi^*(j))) 
%     \leq d(j, \varphi^*(j)) +    d(\varphi^*(j), \varphi(j)) \\
%     &\leq d(j, \varphi^*(j)) + d(\varphi^*(j), j) + d(j, \varphi(j)) 
%     = 2O_j + A_j
%   \end{align*}
% \end{proof}

% \paragraph{Remarks.}
% The main simplification over the paper of Arya et al.~\cite{AGKMMP01}
% comes from the fact that we can define the set of swaps based only on
% the distances between facilities in $F$ and $F^*$, and do not need to
% consider how the 

\subsection{The $k$-Median Local Search with Multiswaps}
\label{sec:kMed-tswaps}

The paper of Arya et al.~\cite{AGKMMP01} also showed that local minima
under $t$-swaps for the $k$-median algorithm have cost at most $(3 +
\frac2t)$ times the optimum, where a $t$-swap involves shutting down $t$
of the current set of $k$ facilities and opening $t$ new facilities in
their stead. (Note that the best $t$-swap can be found in time
$n^{O(t)}$, and hence is polynomial-time for constant $t$.)  Using the
notation from the previous section, we will show that $\kmedcf(F) \leq
(3 + \frac{2}{t})\kmedcf(F^*)$, where $F$ denotes the set of facilities
at the local optimum, and $F^*$ the set of facilities in an optimal
solution.

\subsubsection{A Set of Test $t$-Swaps}
\label{sec:test-tswaps}

In order to extend the proof of the previous section to the $t$-swap
algorithm, we generalize the set of test swaps as follows. The set of
test $t$-swaps relies on the same intuition as before, but it is more
sophisticated for technical reasons. For each element $r$ of $F$,
define the degree of $r$ to be $\deg(r) = |\eta^{-1}(r)|$. The pairing
is constructed by the following procedure.
\begin{algorithm}[H]
  \caption{Pairing construction}
  \label{alg:partition}
  \begin{algorithmic}
    \STATE $i = 0$
    \WHILE {there exists $ r \in F$ such that $\deg(r) > 0$}
    \STATE \texttt{(1)} $R_i = \{r\} \cup \{ \text{any } \deg(r) - 1 
  \text{ elements of } F \text{ with degree } 0\}$. 
    \STATE \texttt{(2)} $F^*_i = \eta^{-1}(R_i)$.
    \STATE \texttt{(3)} 
    $F = F \setminus R_i$, $F^* = F^* \setminus F^*_i$, $i = i + 1$. 
    \ENDWHILE
    \STATE $R_i = F$, $F^*_i = F^*$, $r = i$.
  \end{algorithmic}
\end{algorithm}
% \begin{verbatim}
% [ Will render in an "algorithm" environment later. ]
%    i = 1;
%    while (max degree of f in F > 0)
%       (1) R_i = f \cup { deg(f) - 1 degree-0 elements of F }
%       (2) F*_i = \eta^{-1}(R_i)
%       (3) F = F \ R_i, F^* = F^*\F^*_i
%       (4) i = i + 1;
%    end
%    R_i = F, F^*_i = F^*, r = i.
% \end{verbatim}

When the algorithm terminates, the sequences $\{R_i\}_{i=1}^r$ and
$\{F^*_i\}_{i=1}^r$ form partitions of $F$ and $F^*$, respectively. It
is straightforward to show that \textbf{(a)} in all iterations, step (1)
will be able to find $\deg(r) - 1$ elements of degree $0$, \textbf{(b)}
$|R_i| = |F^*_i|$ for all $i = 1, 2, \dots, r$, and \textbf{(c)} all
elements of $R_r$ have degree $0$. Moreover, the following holds.
\begin{fact}
  \label{fact:nonoverlapping}
  If $j \in R_i$ and $\varphi^*(j) \not\in F^*_i$, then
  $\eta(\varphi^*(j)) \not\in R_i$.
\end{fact}
% \end{obs}
% \end{claim}
% \begin{proof}
%   The proof is immediate from the fact that $F^*_i = \eta^{-1}(R_i)$
%   and that $\{F^*_i\}_{i=1}^r$ forms a partition of the set $F^*$.
% \end{proof}

\subsubsection{The Cost of a Local Optimum: the $t$-Swap Case}
\label{sec:local-opt-tswap}

Based on the potential $t$-swaps, we now give upper bounds on several
possible moves. The following proofs are largely similar to the proof of
\lref[Lemma]{lem:ubd-kmed}.

\begin{lemma}
  \label{lem:ubd-case1}
  If $|R_i| = |F^*_i| \leq t$, then
  \begin{gather}
    \kmedcf( [F \setminus R_i] \cup F^*_i ) - \kmedcf (F)
    \leq \sum_{j \in N^*(F^*_i)} (O_j - A_j) 
    + \sum_{j \in N(R_i)} 2\,O_j
  \end{gather}
\end{lemma}
\begin{proof}
  Consider the following assignment of clients after removing $R_i$
  and adding $F^*_i$: map each client $j \in N^*(F^*_i)$ to
  $\varphi^*(j)$ and map each client $j \in N(R_i)\setminus
  N^*(F^*_i)$ to $\eta(\varphi^*(j))$. All other clients stay where
  they were. We know from \lref[Fact]{fact:nonoverlapping} that this
  assignment is legal, because we do not assign any clients to the
  facilities being removed.  The assignment gives an upper bound on
  the cost of $\kmedcf( [F \setminus R_i] \cup F^*_i )$. Applying the
  projection lemma (\lref[Lemma]{lem:prjlem}), we know that each $j \in
  N(R_i)$ has an increase at most $2O_j$. The rest of the proof is
  similar to the proof of \lref[Lemma]{lem:ubd-kmed}.
\end{proof}

\begin{lemma}
  If $|R_i| = |F^*_i| = s > t$, then
  \begin{gather}
    \frac{1}{s - 1}\smash{\sum_{(f^*,r) \in F^*_i\times \widehat{R}_i}}
    [\kmedcf(F + f^* - r) - \kmedcf(F)] \,\leq\, \sum_{j \in N^*(F^*_i)}
    (O_j - A_j) + \sum_{j \in N^*(R_i)} 2\left(1 + \frac{1}{t}\right)\,O_j,
  \end{gather}
  where $\widehat{R}_i$ is a set of $s - 1$ degree-$0$ elements of $R_i$.
\end{lemma}
\begin{proof}
  Consider each individual swap in $(f^*, r) \in F_i^* \times
  \widehat{R}_i$ that removes $r$ and adds $f^*$.  By
  \lref[Fact]{fact:nonoverlapping}, we know that the assignment in
  \lref[Lemma]{lem:ubd-kmed} is legal. Thus the same analysis as in
  \lref[Lemma]{lem:ubd-kmed} gives that $\kmedcf(F + f^* - r) - \kmedcf(F)
  \leq \sum_{j \in N^*(f^*)} (O_j - A_j) + \sum_{j \in N(r)} 2\,O_j$.

  Now consider each $j \in N^*(F^*_i)$. We know that $f^*$ appears in
  exactly $s - 1$ pairs of $F^*_i\times \widehat{R}_i$, so summing over
  these, we have $\sum_{j \in N^*(F^*_i)} (O_j - A_j)$. Likewise, each
  $j \in N^*(R_i)$ appears in exactly $s$ pairs of $F^*_i\times
  \widehat{R}_i$. But, $\frac{s}{s-1} \leq 1 + \frac1t$, so summing over
  these, we have $\sum_{j \in N^*(R_i)} 2(1+\frac1t)\,O_j$. Together, we
  have the lemma.
\end{proof}

We now combine the two Lemmas. Note that for $|R_i| = |F^*_i| \leq t$,
the optimality condition implies that $\kmedcf([F\setminus R_i] \cup
F^*_i) - \kmedcf(F) \geq 0$, and for $|R_i| = |F^*_i| > t$, the
optimality condition gives that $\kmedcf(F + f^* - r) - \kmedcf(F) \geq
0$ for all $f^* \in F^*_i$ and $r \in R_i$. Therefore, by noting that
$\{R_i\}_{i=1}^r$ and $\{F^*_i\}_{i=1}^r$ are partitions of $F$ and
$F^*$, respectively, we establish
\begin{align*}
  0 &\leq \sum_{i: |R_i| \leq t} \biggl(\kmedcf([F \setminus R_i] \cup F^*_i) - \kmedcf(F)\biggr) + \sum_{i: |R_i| > t} \frac{1}{|R_i|-1}\sum_{(f^*,r) \in F_i^*\times \widehat{R}_i}
  \kmedcf(F + f^* - r) - \kmedcf(F)\\
  &\leq  \sum_{i=1}^r\biggl[ 
    \sum_{j \in N^*(F^*_i)} (O_j - A_j) + \sum_{j \in N(R_i)} 2(1+1/t)\, O_j  \biggr]
    =\kmedcf(F^*) - \kmedcf(F) + 2(1+1/t)\kmedcf(F^*),
\end{align*}
which proves the following theorem.

\begin{theorem}[\cite{AGKMMP01}]
  At a local minimum $F$ for the $t$-swap version of the $k$-median
  local search algorithm, $\kmedcf(F) \leq (3 + 2/t)\,\kmedcf(F^*)$.
\end{theorem}

%%%%%%%%%%%%%%%%%%%%%%%%%%%%%%%%%%%%%%%%%%%%%%%%%%%%%%%%%%%%%%%%%%%%%%%%%%%
% Section: The $\ell_p$-Norm $k$-Facility Location Problem      
%%%%%%%%%%%%%%%%%%%%%%%%%%%%%%%%%%%%%%%%%%%%%%%%%%%%%%%%%%%%%%%%%%%%%%%%%%%
\section{The $\ell_p$-Norm $k$-Facility Location Problem}
\label{sec:lp-kfl}

In this section, we consider the following common generalization of
the $k$-median, $k$-means, and $k$-center problems: given a metric
space with a point set $V$ and distances $d(\cdot,\cdot)$, and a value
$p \geq 1$, the \emph{$\ell_p$-norm $k$-facility location problem} is
to find a set $F$ of $k$ facilities to minimize the objective function
\begin{gather} 
  \label{eq:ellp}
  \ts \Phi_p(F) = \bigg(\sum_{j \in V} d(j, F)^p\bigg)^{1/p}.
\end{gather}

Like the $k$-median problem, this setting has a natural local-search
algorithm: Starting with $k$ facilities, the algorithm tries to find
$i \in F$ and $j\not\in F$ such that the ``swap'' $F' = (F\setminus
\{i\}) \cup \{j\}$ improves the objective value.  The algorithm stops
when no improving move exists.  We will show that this algorithm is a
$5p$ approximation, which can be improved to a $(3+\frac2t)p$
approximation by allowing $t$-swaps.  This algorithm is
\emph{asymptotically tight} as we will show in
\lref[Section]{sec:an-asympt-tight}.  Even for the single-swap case,
this immediately implies a $5$-approximation for $k$-median, a
$10$-approximation for the $k$-means problem (which can be improved to
$9$), and an $O(\log n)$-approximation for the $k$-center problem. The
implication for $k$-center follows from the fact that for vectors of
length $n$, the norm $\norm{\,\cdot\,}{\infty}$ is within a constant
factor of the norm $\norm{\,\cdot\,}{\log n}$, and hence $\max_{j \in
  V} d(j,F)$ is within a constant factor of $\Phi_{\log n}(F)$.  To
the best of our knowledge, the results for $p \neq 1$ give the first
analyses of local-search heuristics for these problems on general
metric spaces.

\subsection{Analyzing the Single-swap Case}

We begin with the single-swap case and analyze the $t$-swap case in
the next section.  Let $\Delta(p, q) = d^p(p,q)$, and assume that $V$
is indexed by $\{1, 2, \ldots, n\}$.  We borrow notations and
definitions from our analysis of $k$-Median in
\lref[Section]{sec:basic-kM}.  To argue about the quality of our
solution, we will use the set $S$ of test swaps defined in
\lref[Section]{sec:test-swaps}.  Using an argument similar to the
proof of the $k$-Median problem, we establish:
\begin{align}
  0 &\leq \sum_{(r,f^*) \in S} \Phi^p_p(F + f^* - r) - \Phi^p_p(F) \leq
  \sum_{(r,f^*) \in S}\left(\sum_{j \in N^*(f^*)} (O^p_j - A^p_j) +
    \sum_{j \in N(r)} \Delta(j, \eta(\varphi^*(j))) - A^p_j \right)\nonumber\\
  &\leq \sum_{j \in V} O^p_j - 3\sum_{j \in V} A^p_j + 2\sum_{j \in V}
  \Delta(j, \eta(\varphi^*(j))) = \Phi^p_p(F^*) - 3\Phi^p_p(F) + 2\sum_{j \in V}
  \Delta(j, \eta(\varphi^*(j))) \label{eq:lpkm-expand}
\end{align}

We proceed to derive an upper-bound on the sum $\sum_{j \in V}
\Delta(j,\eta(\varphi^*(j)))$ in terms of $\Phi_p(F^*)$ and
$\Phi_p(F)$ as follows.
\begin{claim}
  \label{claim:excess-bound}
  \begin{equation*}
    \sum_{j \in V} \Delta(j,\eta(\varphi^*(j))) \leq (2\Phi_p(F^*) +
    \Phi_p(F))^p
  \end{equation*}
\end{claim}

\begin{proof}
  Define $\bm{x} = \bigl\langle d(j, \eta(\varphi^*(j)))
  \bigr\rangle_{j=1}^n$, and $\bm{y} = \bigl\langle 2d(j,\varphi^*(j))
  + d(j,\varphi(j))\bigr\rangle_{j=1}^n$.  The Projection Lemma
  (\lref[Lemma]{lem:prjlem}) gives that $\bm{x}_j \leq \bm{y}_j$ for all
  $j \in V$, and thus $\norm{\bm{x}}{\ell_p} \leq
  \norm{\bm{y}}{\ell_p}$.  Now note that $\Vert\bigvec{d(j,
    \varphi^*(j))}_{j=1}^n\Vert_{\ell_p} = \Phi_p(F^*)$ and
  $\Vert\bigvec{d(j, \varphi(j))}_{j=1}^n\Vert_{\ell_p} = \Phi_p(F)$.
  Therefore, applying the triangle inequality on the $(\R^n,\ell_p)$
  space, we have $ \sum_{j \in V}\Delta(j,\eta(\varphi^*(j))) =
  \norm{\bm{x}}{\ell_p}^p \leq \norm{\bm{y}}{\ell_p}^p \leq
  (2\Phi_p(F^*) + \Phi_p(F))^p$.
\end{proof}

Together with \eqref{eq:lpkm-expand}, this claim implies
\begin{equation}
  0 \leq \Phi_p^p(F^*) - 3\Phi_p^p(F) +2\big(2\Phi_p(F^*) 
    + \Phi_p(F)\big)^p \label{eq:lpkm-goal}
\end{equation}

To complete the proof, we let $\alpha$ be the smallest positive number
such that $\Phi_p^p(F) \leq \alpha^p\cdot \Phi_p^p(F^*)$ and show that
$\alpha \leq 5p$.  Suppose for a contradiction that $\alpha > 5p$.
Consider the right-hand side of inequality \eqref{eq:lpkm-goal}, which
we rewrite as follows.
\begin{align}
  \Phi_p^p(F^*) - 3\Phi_p^p(F) +2\big(2\Phi_p(F^*) + \Phi_p(F)\big)^p
  = \alpha^p\Phi_p^p(F^*)\bigg(\frac{1}{\alpha^p} - 3 + 2(1 + 2/\alpha)^p\bigg)
\end{align}
Let $f(p, \alpha) = \frac{1}{\alpha^p} + 2 (1 + 2/\alpha)^p - 3$.
Since $\alpha^p\Phi_p^p(F^*)$ is always non-negative, if we know that,
for $\alpha > 5p$, $f(\alpha) < 0$, we will have a contradiction to
\eqref{eq:lpkm-goal}.  For $p = 1$ and $2$, we can explicitly solve
for $\alpha$. For $p \geq 2$, we know that if $\alpha > 5p$, then
$\frac{1}{\alpha^p} \leq 1/100$ and $(1+2/\alpha)^p \leq
e^{2/5}$. Thus, we have $\frac{1}{\alpha^p} + 2 (1 + 2/\alpha)^p \leq
\frac{1}{100} + 2\cdot e^{2/5} - 3 < 0$.  The results can be
summarized in the following theorem:

% \medskip

% \textbf{Remarks.} We note that for $p = 2$, the inequality
% \eqref{eq:lpkm-goal} can be solved exactly for $\alpha$. When $p = 2$,
% we have $0 \leq \Phi_p^p(F^*)[1 - 3\alpha^2 + 2(\alpha + 2)^2] =
% \Phi_p^p(F^*)[9 + 8\alpha - \alpha^2]$, which implies that $\alpha
% \leq 9$.

\begin{theorem}
  The natural local-search algorithm for the $\ell_p$-facility
  location problem gives a $5p$ approximation guarantee. Additionally,
  for $p = 2$, this algorithm gives a $9$ approximation guarantee on
  general metric spaces.
\end{theorem}

\subsection{Analyzing the $t$-Swap Case}

We will now analyze the $t$-swaps case.  The analysis here will be
largely similar to the analysis of the single-swap case, except for
few extra ingredients which we will now develop.  Our analysis will
use the set of test swaps from \lref[Section]{sec:kMed-tswaps}.  We will
borrow notations and definitions from our analysis of the $t$-swap
case of $k$-Median in \lref[Section]{sec:kMed-tswaps}.

% Although the proof here will
% be largely similar to the proof for the single-swap case, we will need
% few more ingredients in the $t$-swap case.

% As we have already seen, the natural local-search algorithm for
% the $k$-Median problem can be improved by allowing each swap to
% exchange multiple facilities. In a $t$-swap, $t$ current facilities
% are shut down, and $t$ new facilities are opened in their stead.  In
% this section, we will examine the benefits of multiswaps in the
% $\ell_p$-facility location problem.

% the sequences $\{R_i\}_{i=1}^r$ and
% $\{F^*_i\}_{i=1}^r$

% \paragraph{The Cost of a Local Optimum: the $t$-swap Case.} 

Let us recall that $F$ and $F^*$ denote the facilities in our solution
and the facilities in the optimal solution, respectively; the
sequences $\{R_i\}_{i=1}^r$ and $\{F^*_i\}_{i=1}^r$ denote partitions
of $F$ and $F^*$, respectively.  We derive the following upper bounds.

\begin{lemma}
  \label{lem:lp-ubd-case1}
  If $|R_i| = |F^*_i| \leq t$, then
  \begin{gather}
    \Phi_p^p( [F \setminus R_i] \cup F^*_i ) - \Phi_p^p (F)
    \leq \sum_{j \in N^*(F^*_i)} (O_j^p - A_j^p) 
    + \sum_{j \in N(R_i)} (\Delta(j, \eta(\varphi^*(j))) - A_j^p)
  \end{gather}
\end{lemma}
\begin{proof} 
  Consider the following assignment of clients after removing $R_i$
  and adding $F^*_i$: map each client $j \in N^*(F^*_i)$ to
  $\varphi^*(j)$ and map each client $j \in N(R_i)\setminus
  N^*(F^*_i)$ to $\eta(\varphi^*(j))$. All other clients stay where
  they were. We know from \lref[Fact]{fact:nonoverlapping} that this
  assignment is legal, because we do not assign any clients to the
  facilities being removed.  The assignment gives an upper bound on
  the cost of $\Phi_p^p( [F \setminus R_i] \cup F^*_i )$.  This
  immediately gives $ \Phi_p^p( [F \setminus R_i] \cup F^*_i ) -
  \Phi_p^p (F) \leq \sum_{j \in N^*(F^*_i)} (O_j^p - A_j^p) + \sum_{j
    \in N(R_i)\setminus N^*(F^*_i)} (\Delta(j, \eta(\varphi^*(j))) -
  A_j^p) \leq \Phi_p^p( [F \setminus R_i] \cup F^*_i ) - \Phi_p^p (F)
  \leq \sum_{j \in N^*(F^*_i)} (O_j^p - A_j^p) + \sum_{j \in N(R_i)}
  (\Delta(j, \eta(\varphi^*(j))) - A_j^p)$, which proves the lemma.
% The proof is almost identical to the proof of
%   \lref[Lemma]{lem:ubd-case1}.  Consider the following assignment of
%   clients after removing $R_i$ and adding $F^*_i$: map each client $j
%   \in N^*(F^*_i)$ to $\varphi^*(j)$ and map each client $j \in
%   N(R_i)\setminus N^*(F^*_i)$ to $\eta(\varphi^*(j))$. All other
%   clients stay where they were. We know from
%   \lref[Fact]{fact:nonoverlapping} that this assignment is legal,
%   because we do not assign any clients to the facilities being
%   removed.  The assignment gives an upper bound on the cost of
%   $\Phi_p^p( [F \setminus R_i] \cup F^*_i )$. Applying the projection
%   lemma (\lref[Lemma]{lem:prjlem}), we know that each $j \in N(R_i)$ has
%   an increase at most $O_j$. The rest of the proof is similar to the
%   proof of \lref[Lemma]{lem:ubd-kmed}.
\end{proof}

\begin{lemma}
  \label{lem:lp-ubd-case2}
  If $|R_i| = |F^*_i| = s > t$, then
  \begin{align*}
    \frac{1}{s - 1}\smash{\sum_{(f^*,r) \in F^*_i\times
        \widehat{R}_i}} [\Phi^p_p(F + f^* - r) - \Phi^p_p(F)] &\leq
    \sum_{j \in N^*(F^*_i)} (O_j^p - A_j^p) \\
    &\qquad + \sum_{j \in N^*(R_i)}
    \left(1 + \frac{1}{t}\right)(\Delta(j, \eta(\varphi^*(j))) - A_j^p),
  \end{align*}
  where $\widehat{R}_i$ is a set of $s - 1$ degree-$0$ elements of $R_i$.
\end{lemma}
\begin{proof}
  Consider each individual swap in $(f^*, r) \in F_i^* \times
  \widehat{R}_i$ that removes $r$ and adds $f^*$.  Again, we will
  reassign every $j \in N^*(f^*)$ to $f^*$ and every $j \in
  N(r)\setminus N^*(f^*)$ to $r$, and keep all other clients where
  they were. By \lref[Fact]{fact:nonoverlapping}, we know that this
  assignment is legal, and thus $\Phi^p_p(F + f^* - r) - \Phi^p_p(F)
  \leq \sum_{j \in N^*(f^*)} (O^p_j - A^p_j) + \sum_{j \in N(r)}
  (\Delta(j, \eta(\varphi^*(j))) - A_j^p)$.

  Now consider each $j \in N^*(F^*_i)$. We know that $f^*$ appears in
  exactly $s - 1$ pairs of $F^*_i\times \widehat{R}_i$, so summing over
  these, we have $\sum_{j \in N^*(F^*_i)} (O^p_j - A^p_j)$. Likewise, each
  $j \in N^*(R_i)$ appears in exactly $s$ pairs of $F^*_i\times
  \widehat{R}_i$. But, $\frac{s}{s-1} \leq 1 + \frac1t$, so summing over
  these, we have $\sum_{j \in N^*(R_i)} (1+\frac1t)\,(\Delta(j, \eta(\varphi^*(j))) - A_j^p)$. Together, we
  have the lemma.
\end{proof}

Using these two lemmas and the fact that $\{R_i\}_{i=1}^r$ and
$\{F^*_i\}_{i=1}^r$ are partitions of $F$ and $F^*$, we establish the
following bound:
\begin{align}
  0 &\leq \sum_{i=1}^r\bigg(\sum_{j\in N^*(F^*_i)} (O_j^p -A_j^p) +
  \sum_{j\in N(R_i)} \big(1 + \frac1t\big)(\Delta(j,
  \eta(\varphi^*(j))) - A_j^p) \bigg)\\
  &= \Phi_p^p (F^*) - \big(2 + \frac1t\big)\Phi_p^p(F) +
  \big(1+\frac1t)\sum_{j\in V} \Delta(j, \eta(\varphi^*(j))).
\end{align}

As shown in \lref[Claim]{claim:excess-bound}, the sum $\sum_{j \in V}
\Delta(j,\eta(\varphi^*(j)))$ is upper-bounded by $(2\Phi_p(F^*) +
\Phi_p(F))^p$; therefore, we have
\begin{gather}
  \label{eq:lp-final}
  \Phi_p^p (F^*) - \big(2 + \frac1t\big)\Phi_p^p(F) +
  \big(1+\frac1t)(2\Phi_p(F^*) + \Phi_p(F))^p \geq 0
\end{gather}

We are now ready to prove the following theorem:
\begin{theorem}
  For any value of $t \in \Z_+$, the natural $t$-swap local-search
  algorithm for the $\ell_p$-facility location problem yields the
  following guarantees: 
  \begin{enumerate}

  \item For $p = 1$, it is a $(3 + \frac2t)$ approximation.

  \item For $p = 2$, it is a $(5 + \frac4t)$ approximation.

  \item For all reals $p \geq 2$, it is a $(3 + \frac2t)p$
    approximation.
  \end{enumerate}
\end{theorem}

\begin{proof}
  Let $\alpha$ be the smallest positive number such that $\Phi_p^p(F)
  \leq \alpha^p\cdot \Phi_p^p(F^*)$ in inequality~(\ref{eq:lp-final}).
  For $p = 1$ and $p = 2$, we can directly solve the inequality and
  obtain the desired results. For $p \geq 2$, we assume for a
  contradiction that $\alpha > (3 + \frac2t)p$ and proceed as follows.
  With the assumption, we have
  \begin{align*}
    &\Phi_p^p (F^*) - \big(2 + \frac1t\big)\Phi_p^p(F) +
    \big(1+\frac1t)(2\Phi_p(F^*) + \Phi_p(F))^p \\
    &\qquad\leq \alpha^p\Phi_p^p(F^*)\bigg(\frac1{\alpha^p} - \big(2 +
    \frac1t\big) + \big(1 + \frac1t\big)\big(1 + \frac2\alpha\big)^p \bigg)\\
    &\qquad= \alpha^p\Phi_p^p(F^*)\bigg(\frac1{\alpha^p}+
    \big(1+\frac2\alpha\big)^p - 2 + \frac1t\big(\big(1 +
    \frac2\alpha\big)^p - 1\big)\bigg)
  \end{align*}
  Now define $f(p) = \big(\frac1{(3+2/t)p}\big)^p +
  \big(1+\frac2{(3+2/t)p}\big)^p$ and $g(p) =
  \big(1+\frac2{(3+2/t)p}\big)^p$.  It is easy to see that for $p \geq
  2$, $f(p)$ and $g(p)$ are non-decreasing functions, and thus for all
  $p \geq 2$, $f(p) \leq \lim_{p \to \infty} f(p) =
  \exp\big\{\frac{2t}{2+3t}\big\}$ and $g(p) \leq \lim_{p \to \infty}
  g(p) = \exp\big\{\frac{2t}{2+3t}\big\}$.  Note also that
  $\exp\{\frac{2t}{2+3t}\} = \exp\{1 - \frac{2+t}{2+3t}\} \leq e\cdot
  \big(1 - \frac{2+t}{2+3t} \big) = e\cdot \frac{2t}{2+3t}$.
  \begin{align*}
    &\Phi_p^p (F^*) - \big(2 + \frac1t\big)\Phi_p^p(F) +
    \big(1+\frac1t)(2\Phi_p(F^*) + \Phi_p(F))^p \\
    &\qquad \leq \alpha^p\Phi_p^p(F^*)\bigg(e^{2t/(2+3t)} - 2 +
    \frac1t\big( e^{2t/(2+3t)} - 1\big)\bigg) \\
    &\qquad \leq \alpha^p\Phi_p^p(F^*)\bigg(\frac{1+t}{t}\cdot
    e\cdot\frac{2t}{2+3t} -
    \frac1t - 2\bigg)\\
    &\qquad \leq \alpha^p\Phi_p^p(F^*)\bigg(\frac{2e(1+t)}{2+3t} -
    \frac1t - 2\bigg)
  \end{align*}
  Simple algebra shows that $\frac{2e(1+t)}{2+3t} - \frac1t \leq 1.9$;
  therefore,  $\Phi_p^p (F^*) - \big(2 + \frac1t\big)\Phi_p^p(F) +
  \big(1+\frac1t)(2\Phi_p(F^*) + \Phi_p(F))^p \leq
  \alpha^p\Phi_p^p(F^*)\big(\frac{2e(1+t)}{2+3t} - \frac1t - 2\big)
  < 0$, which gives a contradiction.
% We have inequality~(\ref{eq:lp-final}) can be
%  rewritten as 
%  \begin{equation}
%    
%  \end{equation}
\end{proof}

\subsection{An Asymptotically Tight Example}
\label{sec:an-asympt-tight}
Inspired by a lower bound given by Kanungo et al.~\cite{KMNPSW03}, we
present an example where the local-search algorithm produces an
$\Omega(p)$-approximate solution for the the $\ell_p$-facility
location problem.  The example presented below is designed for the
single-swap case, but it can be generalized to the $t$-swap case. 

\begin{theorem}
  For every $p$, there are instances of the $\ell_p$-norm $k$-facility
  location problem where the cost of some local minima (under the
  standard local moves) is at least $2p$ times the optimal cost.
\end{theorem}

\begin{proof}
  Consider a $2$-dimensional torus with lattice points $\{0, 1, \dots,
  N - 1\}^2$ for some large integer $N$. The lattice points are
  labeled even or odd according to the parity of the sum of their
  coordinates. For a value $x$ to be fixed later, we define a gadget
  $D(x)$ to be a set of $4$ points at $(\pm x, 0), (0,\pm x)$.

  We overlay a graph on the torus as follows. Every lattice point is a
  facility node. Centered at each even lattice point is a copy of
  $D(x)$. These gadget points make up our client nodes. To set up a
  distance metric, we introduce the following edges and define the
  distance between any two nodes as the shortest path between them. As
  shown in \lref[Figure]{fig:pathex1}, an even lattice point is at
  distance $x$ from any of its surrounding gadget points, and an odd
  lattice point is at distance $1 - x$ from its (physical) neighboring
  gadget points. Under this distance metric, the distance between $a$
  and $b$, for example, is $1 + x$.
  \begin{figure}[h!]
    \centering
    \includegraphics[scale=.57]{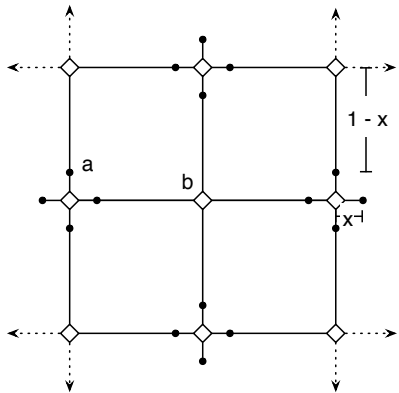}
    \label{fig:pathex1}
    \caption{A lower-bound instance}
  \end{figure}

  Let $x = 1/(2p+1)$ and $k = N^2/2$. For this choice of $k$ and $x$,
  an optimal solution, denoted by $F^*$, opens facilities at all even
  lattice points, yielding $\Phi_p(F^*) = (4k)^{1/p}x$. To get a $2p$
  approximation, consider a solution $F$ which opens $k$ facilities at
  all odd lattice points. It is easy to see that $\Phi_p(F) =
  (4k)^{1/p}(1-x)$, and so $\frac{\Phi_p(F)}{\Phi_p(F^*)} = 2p$. It
  remains to show that $F$ is a local optimum. Consider that, for any
  $f^* \in F$ and $r \in F$, if we shut down a facility $r$ and open a
  facility $f^*$, the change in cost is given by
  \begin{equation}
    \label{eq:lwr-lp}
    \ts \Phi^p_p(F - r + f^*) - \Phi^p_p(F) 
    \geq 4\biggl(x^p - (1-x)^p\biggr) + 3\biggl((1+x)^p - (1-x)^p\biggr).
  \end{equation}
  With our choice of $x$, it is straightforward to show that $
  \Phi^p_p(F - r + f^*) - \Phi^p_p(F) \geq 0$. Thus, the solution $F$ is a
  local optimum as desired.
\end{proof}

\section{Simpler Proofs for Other Previously Known Results}
\label{sec:grab-bag}

\subsection{Uncapacitated Facility Location}
\label{sec:ufl}

In the metric uncapacitated facility location problem, instead of a
hard upper bound $k$ on the number of facilities, we are given an
opening cost $f_i$ for each location $i \in V$, and the goal is to
minimize the objective function 
\[ \uflcf(F) = \sum_{i \in F} f_i +
\sum_{j \in V} d(j, F). \] 

This problem has been extremely widely
studied and many constant-factor approximation algorithms are known: see
\lref[Section]{sec:related-work} for many references.

\paragraph{The Local Search Moves.}
Since we do not have a hard bound on the number of facilities, we can
have a richer set of local moves---\textbf{(a)} opening a facility,
\textbf{(b)} closing a currently open facility, and \textbf{(c)}
swapping facilities as above. Again, we let $F^*$ be the optimal set of
facilities, and $F$ be the algorithm's set of facilities at a local
minimum.

\subsubsection{Bounding the Local Optimum Cost for UFL}

Since we have been using $f$ to denote a generic facility, let us use
$\cost(f)$ to denote the facility opening cost for facility $f$, and
$\cost(F')$ to denote the cost $\sum_{f \in F'} \cost(f)$ of a set $F'$
of facilities. Again, for a client $j \in C$, let $O_j$ and $A_j$ be the
connection cost in the optimal and local-optimal solutions, and let
$\varphi^*$ and $\varphi$ denote the maps assigning clients to
facilities. Hence $\OPT = \cost(F^*) + \sum_{j \in C} O_j$, whereas
$\ALG = \cost(F) + \sum_{j \in C} A_j$. Moreover, for a facility $f \in
F$, let $N(f)$ denote the clients assigned to it; similarly, define
$N^*(f^*)$ to be the clients assigned to it in the optimal solution.
The first lemma below is easy: try opening each facility in $F^*$, note
the change in cost, and add things up.

\begin{lemma}[Connection Cost~\cite{KPR98}]
  \label{lem:kpr-conn}
  At a local optimum, the fact that ``open new facility'' moves are
  non-improving implies the connection cost $\sum_{j \in C} A_j
  \leq \cost(F^*) + \sum_{j \in C} O_j$.
\end{lemma}

\begin{lemma}[Facility Cost~\cite{AGKMMP01}]
  \label{lem:arya-facility}
  The facility cost $\cost(F) \leq \cost(F^*) + 2\sum_{j \in C} O_j$ at
  a local optimum.
\end{lemma}
\begin{proof}
  Recall the notation of \lref[Section]{sec:basic-kM}: given $F^*$ and
  $F$, let $\eta: F^* \to F$ map each optimal facility to a closest
  facility in $F$.  Following~\cite{AGKMMP01}, call a facility $f \in F$
  ``good'' if $\eta^{-1}(f) = \emptyset$, and ``bad''
  otherwise. 

  If a facility $f$ is good, we can consider closing the facility and
  assigning any clients $j \in N(f)$ to $\fhat = \eta(\varphi^*(j))$:
  note that since $f$ is good, we know that $\fhat \neq f$, and hence
  this reassignment is valid. By the projection lemma
  (\lref[Lemma]{lem:prjlem}), the total increase in the assignment cost
  is $d(j, \fhat) - d(j, f) \leq 2\,O_j$,
%   \begin{align}
%     d(j, \fhat) - d(j, f) &\leq 
%        d(j, \varphi^*(j)) + d(\varphi^*(j),  \fhat) - d(j,f) \\
%     &\leq d(j, \varphi^*(j)) + d(\varphi^*(j), f) - d(j,f) \leq 2\,d(j,
%        \varphi^*(j)) = 2\,O_j \label{eq:5},
%   \end{align}
  and hence from local optimality, we get that for any good $f \in F$,
  \begin{gather}
    - \cost(f) + \sum_{j \in N(f)} 2\,O_j \geq 0. \label{eq:11}
  \end{gather}
  
  For a bad facility $f$, let $P^*_f$ be the set $\eta^{-1}(f) = \{\g_0,
  \g_1, \ldots, \g_t\}$ (with $t \geq 0$), and let $\g_0$ be the closest
  one to $f$. We then consider the $t$ possible moves of opening
  facility $\g_i$ in $P^*_f \setminus \{\g_0\}$, and assigning any client $j
  \in N^*(\g_i) \cap N(f)$ to $\g_i$. The local optimality ensures that
  \begin{gather}
    \cost(\g_i) + \sum_{j \in N^*(\g_i) \cap N(f)} (O_j - A_j) \geq 0.
    \label{eq:7}
  \end{gather}
  
  Moreover, consider the move of opening $\g_0$ and closing $f$: 
  \begin{itemize}
  \item Any client $j \in N(f)$ with $\varphi^*(j) \not \in P^*_f$ is
    assigned to the facility $\eta(\varphi^*(j)) \neq f$: the projection
    \lref[Lemma]{lem:prjlem} implies that the increase in connection cost for such $j$ is
    at most $2\, O_j$.
    
  \item Any client $j \in N(f)$ with $\varphi^*(j) = \g_i \in P^*_f$ (for some
    $i \in \{0, 1, \ldots, t\}$ is assigned to $\g_0$. The change in the
    connection cost is $d(j, \g_0) - d(j, f)$. 
  \end{itemize}
  Hence, local optimality shows that
  \begin{gather}
    \cost(\g_0) - \cost(f) + \sum_{j \in N(f) \land \varphi^*(j) \not\in P^*_f}
    2\, O_j + \sum_{i = 0}^t \sum_{j \in N^*(\g_i) \cap N(f)} (d(j, \g_0)
    - A_j) \geq 0. \label{eq:6}
  \end{gather}
  Adding~(\ref{eq:6}) with the $t$ inequalities~(\ref{eq:7}) (one for
  each $i \in \{1, \ldots, t\}$) gives us
  \begin{gather}
    \cost(P^*_f) - \cost(f) + \sum_{j \in N(f) \land \varphi^*(j) \not\in P^*_f}
    2\, O_j + \sum_{i = 0}^t \sum_{j \in N^*(\g_i) \cap N(f)} (d(j, \g_0)
    + O_j - 2\,A_j) \geq 0. \label{eq:8}
  \end{gather}
  Consider the rightmost sum in~(\ref{eq:8}): for $i = 0$, the summand is $2(O_j - A_j)
  \leq 2\,O_j$. For $i \neq 0$, 
  \begin{align}
    d(j, \g_0) + d(j, \g_i) - 2\,d(j, f) 
    &\leq (d(j, f) + d(f, \g_0)) + d(j, \g_i) - 2\,d(j,f) \\
    &\leq d(f, \g_i) + d(j, \g_i) - d(j,f) \label{eq:9} \\
    &\leq 2\,d(j, \g_i) = 2\,O_j,
  \end{align}
  where we used the fact that $d(f,\g_0) \leq d(f,\g_i)$ in~(\ref{eq:9}),
  and the triangle inequality at other places. Now the
  expression~(\ref{eq:8}) can be simplified to say
  \begin{gather}
    \ts \cost(P^*_f) - \cost(f) + \sum_{j \in N(f)} 2\, O_j \geq 0.
    \label{eq:10}
  \end{gather}
  Summing~(\ref{eq:10}) over all bad $f$, and~(\ref{eq:11}) over all the
  good $f$, we get
  \begin{gather}
    \ts \cost(F^*) - \cost(F) + \sum_{j \in C} 2\,O_j \geq 0.
  \end{gather}
  which proves the claimed bound $\cost(F) \leq \cost(F^*) + \sum_j 2\, O_j$.
\end{proof}

%\ktnote{Do we want to have a summarizing thm?}

Combining the facility cost and connection cost lemmas above results
in the following theorem.
\begin{theorem}
  At a local optimum, $\uflcf(F) \leq 2\cost(F^*) + 3\,\sum_j O_j \leq
  3\,\uflcf(F^*)$.
\end{theorem}

\subsection{$k$-Uncapacitated Facility Location}

Building on the techniques developed in the previous sections, we can
now give proofs for the \emph{metric $k$-uncapacitated facility location
  problem} ($k$-UFL) problem. This is a common generalization of the
$k$-median and UFL problems: not only do we have an opening cost $f_i$
for each location (like UFL), but we also have a limit $k$ on the number
of facilities (like $k$-median).  The goal is still to minimize the cost
\[ \kufl(F) = \cost(F) + \sum_{j \in V} d(j, F), \] where $\cost(F) =
\sum_{i \in F} f_i$. This problem was defined by~\cite{DGKPSV05}, whose
main result---showing that local search was a $5$-approximation---is
reproved in this section.

\paragraph{The Local Search Moves.} 
We start with any set $F$ of at most $k$ facilities, and allow the
following modes: the algorithm can open a new facility if $|F| < k$, it
can close a facility in $F$, or swap an open facility in $F$ with a
currently closed facility; as usual, the algorithm performs a move only
if the total cost decreases.

Since we want to argue about some local optimum $F$, we can assume that
$|F| = k$: indeed, if $|F| < k$, then it is a local optimum with respect
to all moves---opening, closing, or swapping facilities, and then the
result of \lref[Section]{sec:ufl} shows that $F$ is a $3$-approximation.
However, if $|F| = k$, then we cannot open facilities even if we want
to, and hence have to work harder for the proof.

\subsubsection{Pairing for $k$-UFL}

As with previous proofs, the central ingredient of the proof is an
appropriate pairing that allows us to bound the cost of the local-search
solution by applying the local optimality condition. To generate the
pairing, we proceed as follows.
\begin{itemize}
\itemsep=1pt
\item Pair degree-$1$ facilities $f \in F$ with $\eta^{-1}(f)$. The
  degree-$1$ group is called ``single.''

\item Pair higher degree facilities in the following manner. If
  $\deg(f) \geq 2$, let $P^*_{f^*} = \eta^{-1}(f) = \{f^*_0, f^*_1,
  \dots, f^*_t\}$, where $f^*_0$ is the facility closest to $f$, and
  the facilities $f^*_1, \dots, f^*_t$ are ordered
  arbitrarily. Additionally, let $f_0 = f$ and $f_1, \dots, f_t$ be
  any $t$ distinct degree-$0$ facilities in $F$. Since $|F| \leq |F^*|$, we will
  always be able to find enough degree-$0$ facilities. Let $P_f =
  \{f_0, f_1, \dots, f_t\}$. The pairs $(P_f, P^*_{f^*})$ are called
  ``heavy strips.''

\item At this point, some degree-$0$ facilities are still left
  unmentioned; we call them ``excess'' facilities.
\end{itemize}

\subsection{Bounding the Cost of a Local Optimum}

We apply the local optimality condition to the pairs as follows.  For
each single pair $(f, f^*)$, we could swap $f^*$ for $f$, assigning
all $j \in N^*(f^*)$ to $f^*$ and $j \in N(f)\setminus N^*(f^*)$ to
$\eta(\varphi^*(j))$.  The same reasoning as before shows that this is
a valid reassignment. The local optimality condition, together with
the projection \lref[Lemma]{lem:prjlem}, gives
\begin{gather}
  \cost(f^*) - \cost(f) + \sum_{j \in N^*(f^*)} (O_j -A_j) 
    + \sum_{j \in N(f)\setminus N^*(f^*)} 2O_j \geq 0
    \label{eq:case0}
\end{gather}

Now consider a heavy strip $(P_f, P^*_{f^*})$. Suppose $P_f = \{f_0,
\dots, f_t\}$ and $P^*_{f^*} = \{f^*_0, \dots, f^*_t\}$. First, we
could swap $f^*_0$ for $f_0$, assigning all $j \in N^*(f^*_0)$ to
$f^*_0$, all $j \in N^*(f^*_i) \cap N(f_0)$ (for $i = 1,2, \dots, t$)
to $f^*_0$, and all $j \in N(f_0)\setminus (\cup_{i=1}^t N^*(f^*_i))$
to $\eta(\varphi^*(j))$.  This is a legal assignment as can be easily
checked. By the local optimality condition, we have
\begin{align}
  &\cost(f^*_0) - \cost(f_0) 
  + \sum_{j \in N^*(f^*_0)} (O_j - A_j) \nonumber\\ 
  &\quad + 
  \sum_{i=1}^t \smash{\sum_{j \in N^*(f^*_i) \cap N(f_0)}} (d(j, f^*_0) - A_j) 
   +
  \sum_{j \in N(f_0)\setminus (\cup_{i=1}^t N^*(f^*_i))} 2O_j \geq 0
  \label{eq:case1}
\end{align}
Within the same strip $(P_f, P^*_{f^*})$, we can also exchange $f_i$
for $f^*_i$ (for each $i = 1, 2, \dots, t$):
\begin{itemize}

\item If we assign all $j \in N^*(f^*_i) \cap (N(f_0) \cup N(f_i))$ to
  $f^*_i$ and all $j \in N(f_i)\setminus N^*(f^*_i)$ to
  $\eta(\varphi^*(j))$, then the local optimality condition yields
  \begin{gather}
    \cost(f^*_i) - \cost(f_i) + \sum_{j \in N^*(f^*_i) \cap (N(f_0) \cup N(f_i))} 
    (O_j - A_j)
    + \sum_{j \in N(f_i)\setminus N^*(f^*_i)} 2O_j \geq 0.
    \label{eq:case2}
  \end{gather}

\item If we assign all $j \in N^*(f^*_i)$ to $f^*_i$ and all remaining
  $j \in N(f_i)$ to $\eta(\varphi^*(j))$, then the local optimality
  condition yields
  \begin{gather}
    \cost(f^*_i) - \cost(f_i) + \sum_{j \in N^*(f^*_i)} (O_j - A_j) +
    \sum_{j \in N(f_i)\setminus N^*(f^*_i)} 2O_j \geq 0.
    \label{eq:case3}
  \end{gather}

\end{itemize}

Finally, consider deleting the ``excess'' facilities. For each excess
facility $f$, we could delete it and assign all facilities $j \in
N(f)$ to $\eta(\varphi^*(j))$---recall $f$ has degree $0$. By the
local optimality, we have
\begin{gather}
  - \cost(f) + \sum_{j \in N(f)} 2O_j \geq 0.
  \label{eq:excesscase}
\end{gather}

Adding up \eqref{eq:case0}-\eqref{eq:excesscase} across all strips and
all groups yields the following claim.
\begin{claim}
  For any $j \in C$, the increase in connection cost is upper bounded
  by $5O_j - A_j$.
\end{claim}

\begin{proof} Consider a client $j$. Note that $j$ is uniquely
  assigned to a facility $f^* = \varphi^*(j)$ in the optimal solution.
  If $f^*$ is in the degree-$1$ group, the increase in connection cost
  is clearly upper bounded by $5O_j - A_j$.  Otherwise, $f^*$ belongs
  to a heavy strip $(\{f_0, \dots, f_t\},\{f^*_0, \dots, f^*_t\})$, in
  which case we consider the following possibilities: \vspace{-2mm}
  \begin{itemize}
    \itemsep=0pt
  \item If $j \in N^*(f_0)$, then the increase in connection cost is
    upper bounded by $2O_j + 2O_j + (O_j - A_j) \leq 5O_j - A_j$.
    
  \item If $j \in N^*(f^*_i) \cap N(f_0)$ for some $i \in [t]$, then
    the increase in connection cost is upper bounded by $d(j, f_0^*) -
    A_j+ (O_j - A_j) + (O_j - A_j) \leq 3O_j - A_j$. The inequality
    follows from the fact that $d(j, f_0^*) \leq d(j, \varphi(j)) +
    d(\varphi(j), f_0^*) \leq d(j, \varphi(j)) + d(\varphi(j), f_i^*)
    \leq d(j, \varphi(j)) + d(\varphi(j), j) + d(j,f^*_i) = 2A_j +
    O_j$. 

  \item If $j \in N^*(f^*_i) \cap N(f_i)$ for some $i \in [t]$, then
    the increase in connection cost is clearly upper bounded by $2(O_j
    - A_j) \leq 5O_j - A_j$.

  \item Otherwise $j$ must belong to $N^*(f^*_i) \setminus (N(f_0)
    \cup N(f_i))$. In this case, the increase in connection cost is
    upper bounded by $2O_j + 2O_j + O_j - A_j = 5O_j - A_j$.
  \end{itemize}\vspace{-2mm}
  We conclude that for all $j \in C$, the increase in connection cost is
  at most $5O_j - A_j$.
\end{proof}

It follows that $2\cost(F^*) - \cost(F) + \sum_{j \in C} (5O_j - A_j)
\geq 0$, resulting in the following theorem.
\begin{theorem}
  At a local minimum, $\kufl(F) \leq 2\cost(F^*) + 5\sum_j d(j,F^*) \leq
  5\, \kufl(F^*)$.
\end{theorem}

{\footnotesize \bibliography{abbrev,embedding}}
%{\small \bibliography{../abbrev,../embedding,../my-papers}}
%{\small \bibliography{../../abbrev,../../embedding,../../my-papers,../robustloc}}
%{\small \bibliography{../embedding}}
\bibliographystyle{alpha}

\appendix

% If we add \eqref{eq:case1} with inequalities \eqref{eq:case2} and
% \eqref{eq:case3}, one for each $i = 1, \dots, t$, we have the
% following inequality:
% \begin{align}
%   &2\cdot\cost(P^*_{f^*}) - \cost(P_f) +
%   \sum_{j \in N^*(f^*_0)} (O_j - A_j) +
%   \sum_{i=1}^t \sum_{j\in N^*(f^*_i)\cap N(f_0)}(d(j,f^*_0) - A_j + 2[O_j - A_j]) 
%   \nonumber \\
%   &\quad  + \sum_{i=1}^t\sum_{j \in N^*(f^*_i)\cap N(f_i)} (O_j - A_j + O_j) 
%   + \sum_{i=1}^t\sum_{j \in N^*(f^*_i)\setminus (N(f_0) \cup N(f_i))} 
%   (4O_j + O_j - A_j) \geq 0
% \end{align}

% It can be shown using simple applications of triangle inequality that
% $d(j, f^*_0) - A_j \leq O_j + A_j$, so $d(j,f^*_0) - A_j + 2(O_j -
% A_j) \leq 3O_j - A_j$. From the expression above, we know that for all
% $j \in N(P_f) \cup N^*(P^*_{f^*})$, the increase is upper bounded by
% $5O_j - A_j$.

% Summing up all three groups, we have that
% \begin{align*}
%   2\cost(F^*) - \cost(F) + \sum_{j \in C} (5O_j - A_j) \geq 0,
% \end{align*}
% which yields the following theorem.

% \begin{theorem}
%   At a local minimum, $\cost(F) + \Phi(F) \leq 2\cost(F^*) +
%   5\Phi(F^*)$.
% \end{theorem}

\end{document}